\documentstyle[aps,preprint]{revtex}
\input psfig
\begin{document}
\draft
\preprint{OSU-TA-21/95}
\date{October 19, 1995; revised February 7 1996}
\title{Phase Transitions at Preheating}

\author{I. I. Tkachev}
\address{Department of Physics, The Ohio State University,
Columbus, OH 43210\\ 
and \\
Institute for Nuclear Research of the Academy of Sciences of Russia
\\Moscow 117312, Russia}

\maketitle

\begin{abstract}
Symmetry restoration processes during the non-equilibrium stage of 
``preheating'' after inflation is studied. It is shown that symmetry 
restoration is very efficient when the majority of created particles are 
concentrated at energies much smaller than the temperature  $T$ in equilibrium. 
The strength of symmetry restoration measured
in terms of the equivalent temperature can exceed $T$ by many orders
of magnitude. In some
models the effect can be equivalent to that if the temperature of instant 
reheating would be close to the Planck scale. This can have an important 
impact on GUT and axion models.
\end{abstract}


%\newpage
\narrowtext
In accordance to modern cosmology 
the outcome of the earliest stages of the Universe evolution, which predefines
its modern appearance, is determined by fine details of the dynamics 
of a particular scalar field, which is called inflaton. For a review of inflationary 
models see  Ref. \cite{al90}. The chaotic inflation \cite{al83} is
a typical model which has essential common features. In this model the inflationary 
stage itself persists till slowly decreasing inflaton field, $\phi (t)$, 
is larger than the Plank scale, $M_{\rm Pl} \approx 10^{19}$ GeV. 
During this stage the Universe expands exponentially and the room for the future matter
is created. This stage ends when the inflaton field reaches $\phi \sim M_{\rm Pl}$ 
and then the field starts to oscillate
coherently. Coherently oscillating field can be considered as a collection
of unstable inflaton quanta at rest, so it decays to all particles it has
coupling with, and the matter is created.

With the assumption of ``instant'' reheating the products of inflaton decay
would thermalize on time scale negligible compared to the rate of the Universe
expansion, so the temperature after reheating would be $T^4_{\rm eq} = 
\rho_\phi/g_*$, where numerical factor $g_*$ includes total number of degrees
of freedom and is large, $g_* \agt 10^2$. Here $\rho_\phi$ is the initial energy 
density stored in inflaton oscillations $\rho_\phi \sim  \lambda \phi^4/4 \sim 
 \lambda M_{\rm Pl}^4$. It is relatively low since the inflaton self-coupling
constant has to be very small $\lambda \sim 10^{-13}$ for the induced density
perturbations to satisfy observational constraints.
In reality, the reheating temperature
will be even much smaller since the reheating is not instant and while the
particles thermalize, the Universe expands and cools.
 
The magnitude of the reheating temperature after inflation is considered to be
important as, for example, this will determine  whether or not certain scenarios
of baryogenesis in Grand Unified Theories (GUT) of strong and elecktroweak interactions
will be successful which requires the reheating to be up to the GUT scale 
$M_X \sim 10^{16}$ GeV. Another important issue is the
occurrence of phase transitions. If the reheating temperature is larger
than the Grand Unification  symmetry 
breaking scale, then the corresponding symmetry will be restored 
(for reviews of phase transitions in GUT, see, 
e.g. \cite{kst87,al90}). The subsequent cooling
will be accompanied by a symmetry breaking phase transition which will
proceed in different horizon volumes independently, resulting in
creation of topological defects: domain walls, strings and magnetic monopoles.
This means that the problem of monopoles \cite{zh} and domain walls \cite{zko} 
will be resurrected, ruling out the corresponding models. 

The question of symmetry restoration is interesting not only in connection
to topological defects.
Another important aspect is whether the Peccei-Quinn (PQ) symmetry \cite{pq} is 
restored after inflation, or not. PQ
symmetry was introduced to explain the apparent smallness of 
CP-violation in QCD. Pseudo-Nambu--Goldstone boson resulting from the 
spontaneous breaking of this symmetry, the invisible axion, is among the best 
motivated candidates for cosmic dark matter. The combination of cosmological and
astrophysical considerations restrict the relevant symmetry breaking scale,
or axion decay constant $f_a$, to be in the narrow window $10^{10}\, {\rm
GeV} \leq f_a \leq 10^{12}\, {\rm GeV}$ (for a review see \cite{t90},
note that the upper bound on $f_a$ does not apply in certain
inflationary models \cite {al91}).
If the PQ symmetry is restored after inflation, then 
the axion field will not be constant throughout the
Universe, but will
have independent values in different horizons. These fluctuations in the
axion field are transformed into density fluctuations of order
unity at the crucial epoch when the axion mass switches on at $T \approx 1$ 
GeV, leading to the existence of very dense axion 
miniclusters \cite{am,kt93}, which may be observable \cite{kt95}.
This also shifts the main source of axions from a coherent
misalignment angle to decaying axion strings \cite{axs}. 

Traditionally, the answer to all of the above questions was associated with the value
of the reheating temperature after inflation.
The purpose of the present Letter is to show that the reheating temperature
is actually irrelevant here and all processes of interest are even
more efficient while the system is still out of equilibrium. This is
more or less apparent for the baryogenesis since one of the necessary
conditions for baryon number generation is the absence of
equilibrium. We shall show that symmetry restoration is more efficient
too in a non-equilibrium state generated at the final stage of inflation.

Recently it was realized that in certain cases the decay of the inflaton
field can be a very fast process \cite{kls94,str,b95} (see also \cite{fid}), 
owing to the possibility
of stimulated decays. This is also called the parametric resonance, for the
general theory of it see, e.g. \cite{gmm}. Sometimes the parametric 
resonance is considered as a very special type of decay of coherent field which
can not be described as a decay of particles at rest. This is an incorrect conclusion -
one just has to include stimulated processes in addition to spontaneous 
\cite{it87} for the particles decay or annihilation. 
That  is why parametric resonance exists only for Bosons in 
the final state. 
But even in Bose systems the parametric resonance is not necessarily always
effective. For example, it was studied long ago for the decay of the axion field
\cite{daf,it87} with the negative conclusion that the expansion of the universe
removes particles from the narrow resonance zone too quickly, blocking the 
entire process
(to reach stimulated decays of axions bound in a gravitational well is not
impossible in principle, but would require enormous densities of particles in this
particular case \cite{it87,kw}). This means, in particular, that if the inflaton 
potential and inflaton interactions are constructed in analogy with 
the axion potential, which is the case in the model of ``natural'' inflation 
\cite{ni}, then the parametric resonance can be ineffective. However, as has 
been demonstrated in \cite{kls94}, a successful parametric resonance can occur in 
the case of, e.g., chaotic inflation  \cite{al83}.

In the case of successful resonance almost all energy stored in the form of 
coherent inflaton oscillations is transferred almost instantaneously to radiation, 
but the  products of inflaton decay
are still far from equilibrium. This intermediate stage was 
dubbed ``preheating" in Ref. \cite{kls94}.
To simplify subsequent discussion
let us first take the  distribution function 
of created particles to be of the form
\begin{equation}
f(p)=A \delta (p_0 - E)
\label{e1}
\end{equation} 
For the case of two particle decay $E$ in this equation 
is equal to half the inflaton mass, for $2 \rightarrow 2$ annihilation
$E =m_\phi$, for the processes $4 \rightarrow 2$ of self-annihilation 
we have $E =2m_\phi$, etc. The main point which is crucial for the subsequent 
discussion is that $E$ is typically smaller by many orders of 
magnitude than the temperature $T$ of instant reheating. Equilibration will
take time, meanwhile particular phase transitions can occur in the system 
when it is still far from the equilibrium and the distribution function is still 
given by Eq. (\ref{e1}). Moreover, we shall show that in this case the symmetry 
restoration is even much more efficient.

Before we proceed, let us specify the general field and particle contents of the system.
First, there is classical inflaton field which we already introduced as $\phi$.
We denote as $m_\phi$ the effective inflaton mass at the Plank scale, which 
includes contribution from the self-interaction
$m_\phi^2(\phi) = m_\phi(0)^2 + 3\lambda\phi^2$. For us it is only important that
the overall value of $m_\phi$ is fixed to $m_\phi \sim 10^{13}$ GeV by the observed 
magnitude of density perturbations.
Second, there are products of inflaton decay. We denote them as $\eta$ and
the mass of corresponding quanta is $m_\eta$. 
The possibility of the inflaton decay into $\eta$-quanta assumes that there
is interaction of, say, the form $g\eta^2\phi^2/2$. In reality, there can
be many channels for the inflaton decay and the final answer is the sum over
all species. Since the true content is unknown, we shall not carry out this
summation, but implicitly assume that it has to be done. Third, there is
relevant order parameter, the classical field $\Phi$. The $\eta$-particles couple 
to the order parameter, so that their mass depends upon it, 
$m_\eta = m_\eta(\Phi)$. The typical case is 
$m_\eta^2(\Phi) = m_\eta(0)^2 + \alpha\Phi^2$ 
with $\alpha$ being a product of coupling constant and some numerical 
factor which depends upon particular direction in the space of internal 
symmetries. In the case of simple one-component scalars this assumes interaction
of the form $\alpha\eta^2\Phi^2/2$. 

Note, that since we have to carry out the summation over all channels, there will 
be terms when $\eta$ corresponds to quanta of the $\phi$ or $\Phi$ fields, and 
even when $\phi$, $\eta$ and $\Phi$ is one and the same field. In some models some 
particular terms can be 
negligible. The simplest possibility corresponds to the only one dominant term 
$\phi=\eta=\Phi$. While we do not exclude this possibility, in order to keep 
uniformity we shall keep separate notations for the separate aspects of one and 
the same entity. For example, we shall still denote by $\eta$ the quanta of the 
$\phi$ field, etc.

In the vacuum state the symmetry is broken, which can be described at the tree level by
the potential $V_0 = -\mu^2 \Phi^2/2 + \lambda_\Phi \Phi^4/4$.
This relates $\mu$ to the symmetry breaking scale via $\mu^2 =\lambda_\Phi \Phi^2$.
At non-zero density of particles there are corrections to this potential.
In particular, $m^2_\Phi (0) \equiv d^2V_0/d\Phi^2 \, (\Phi =0)$ became positive and 
the symmetry is restored. 
To calculate the modified potential we can make use of the fact that the
effective potential is minus the pressure. This definition is 
physically transparent:
if two phases can coexist, the phase with the lower value of the 
effective potential will have higher pressure and the bubbles of 
this phase will eventually occupy the whole volume.

The pressure $P$ for an arbitrary distribution function of $\eta$-particles can be found 
by using the formula
\begin{equation}
T^{\mu \nu} = \int \frac{d^3p}{(2 \pi)^3} \frac{p^\mu p^\nu}{p^0} f(p)  \, ,
\label{e2}
\end{equation}
where $p_0^2 =p^2 + m_\eta^2 (\Phi)$ and for an isotropic medium we have 
$T^{ij} =\delta^{ij} P$.  We omit the subscript $\eta$ for the distribution 
function since $\eta$ are the only particles we have.

While the field $\Phi$ evolves in the effective potential, the number of 
particles does not change on time scales of interest, so
we shall calculate (\ref{e2}) assuming that
\begin{equation}
N = \int \frac{d^3p}{(2 \pi)^3} f(p)  \, ,
\label{e4}
\end{equation}
is constant which fixes the normalization factor $A$ in Eq. (\ref{e1}). 
The procedure is very simple and the result 
for the particle-dependent part of the effective potential is 
$V_1(\Phi) = -P = N\, [m_\eta^2(\Phi) -E^2]/3E$.  Since for the distribution function 
given by Eq. (\ref{e1}) the energy density is simply $\rho =N E$, and 
since with the assumption of instant reheating
the energy density would be unchanging, it is convenient to rewrite our  
result in the form 
\begin{equation}
V_1(\Phi) = \frac{\rho}{3 E^2}m_\eta^2(\Phi) - \frac{\rho}{3} \, .
\label{e5}
\end{equation}

At $m_\eta^2=0$ we recover the equation of state for massless particles, $P= \rho/3$. 
But $\rho$ is a $\Phi$-independent constant and is insignificant for us here. 
What we are interested in is the coefficient
in front of $m_\eta^2(\Phi)$, which we denote $B$, 
\begin{equation}
B =\frac{\rho}{3 E^2}. 
\label{Bd}
\end{equation}
This coefficient is positive, so when added to the negative mass square
of the field $\Phi$ in vacuum, 
leads to the restoration of the symmetry at large density of particles,
$V^{\prime \prime}(0)=-\mu^2 + 2\alpha B$. The symmetry is restored at 
$B >\mu^2/2\alpha \sim (\lambda_\Phi/\alpha)\Phi^2$ (in the
typical case of positive $\alpha$; negative $\alpha$ is also possible
in models with several scalar fields, 
see Ref. \cite{kst87}, which breaks the symmetry instead).

If we were to calculate the effective potential with an equilibrium
thermal distribution we would obtain $B_{\rm eq} = T^2 /24$, see Ref. \cite{1l}. 
Roughly, in Eq. (\ref{Bd}), we would have in this case $\rho \sim T^4$ and $E \sim T$.  
We can generalize
the expression for $B$ as been given by the ratio of particle density to 
the mean energy of particles for a case when the width of the distribution
is finite, but still it is smaller than $E$.

Now we can compare the effectiveness of the symmetry restoration at preheating 
to the instant reheating (actual reheating in the expanding Universe is even less effective). 
In both cases the energy density is the same and is equal to initial inflaton energy 
density, but at preheating $E \ll T$. The restoration
of the symmetry is much more effective in the non-equilibrium state Eq. (\ref{e1}), 
its strength amplified by $B \sim (T/E)^2 B_{\rm eq}$.

Let us make an order of magnitude estimate for $B$ in some possible inflationary models.
The inflaton field strength at the end of inflation is of order $M_{\rm Pl}$, so that 
the energy density in inflaton oscillations is given by $\rho \sim m_\phi^2 M_{\rm Pl}^2$. 
Its magnitude is fixed by the magnitude
of primordial density perturbations. To use Eq. (\ref{Bd}) we only need the estimation 
of $E$. We can consider three cases here.

1) If $E \sim m_\phi$ we obtain
$B \sim  M_{\rm Pl}^2$ , which is equivalent to a would-be reheating temperature up to 
the Plank scale. Note, however, that we had not calculated the numerical coefficient here, 
which can be rather small. One example when this regime can be valid corresponds to 
the self-annihilation of the inflaton field, i.e. $\eta=\phi$ dominated case. 
Decay corresponding to the  $4 \rightarrow 2$ processes
is unsuppressed in the expanding universe if the inflaton mass is dominated by 
the self-interaction, and $E \approx 2m_\phi(\phi)$ \cite{kls94}.

This effect will cause the inflaton field to ``roll back'' to some extent in the new 
inflationary scenario 
(the inflaton can not roll all the way back to the origin if $\phi \approx 0$ was the
initial condition), which was observed in the detailed numerical simulations of 
Ref. \cite{b95}. 

2) Let us consider the axion case (or to this extent any model with sufficiently low 
value of symmetry breaking scale). The Peccei-Quinn symmetry breaking 
scale $f_a \alt 10^{12}$ GeV is comparable to the inflaton mass. Therefore the 
contribution to the mass of the $\eta$ particles due to interaction with PQ field is smaller
than the inflaton mass and no additional kinematical constraints appear.
However, the inflaton will decay into lowest zones where $E \sim m_\phi$
only if $g \sim \lambda$. This happens at $g \ll \lambda$ too, but then the dominant
process is self-annihilation of the inflaton to its own quanta.
For the case $g \agt \lambda$ the mean energy of created particles was found in 
Ref. \cite{kls94} to be given by $\bar{E}^2 \sim g^{1/2} m_\phi M_{\rm Pl}$ and
we find $B \sim g^{-1/2} m_\phi M_{\rm Pl} \sim (\lambda/g)^{1/2} M_{\rm Pl}^2$.
Nevertheless, this might not reduce the coefficient $B$ significantly since
in a models without special symmetry cancellations the constant $g$ can not be very large, 
otherwise loop corrections will induce unacceptably large self-coupling for the inflaton.
In such models we have $g \alt \sqrt{\lambda} $. 
Moreover, even assuming $g \alt 1$, for the effective equivalent temperature
which would result to the same strength of the symmetry restoration we find  
$T_{\rm eff} \agt (\lambda)^{1/4} M_{\rm Pl} \sim 10^{-3} M_{\rm Pl}$.

The Peccei-Quinn symmetry is restored if 
$B \agt (\lambda_\Phi/\alpha)f_a^2 \sim 10^{-14}(\lambda_\Phi/\alpha)M_{\rm Pl}^2$.
We see that the PQ symmetry is restored at preheating if
$\alpha >10^{-7}\lambda_\Phi \sqrt{g}$, which is a rather weak condition. 
Note also that in chaotic inflationary model the PQ symmetry can be restored anyway 
already during inflationary stage, just due to possible direct coupling of inflaton
and PQ fields \cite {kla}.  
We conclude that the ``thermal'' scenario for the axion evolution has strong support.

3) Now let us consider the case of GUT which has large magnitude of the symmetry
breaking scale $\Phi \sim 10^{16}$ GeV. Since $E$ can not be smaller than $m_\eta$, 
the parameter $B$ is suppressed for $\eta$ particles which have large coupling to $\Phi$, 
and which correspondingly 
have large masses $M_X$. However, at fixed coupling $g$
the decays to instability zones with large energy are suppressed, and we 
possibly can neglect those channels even if the value of $g$ is large.
In Ref. \cite{kls94} it was shown that the creation rate of particles does not
depend much upon $g$ near the surface of zero energy of created particles.
Consequently the inflaton will preferably decay to particles which have low value of 
$\alpha$. Since the particle content of the theory is large, we can expect that 
particles satisfying the condition 
$m_\eta^2 \sim \alpha\Phi^2  < g^{1/2}m_\phi M_{\rm Pl}^2$ when there
is no additional suppression, can be found. For the case of GUT's this translates
to the condition $\alpha \alt \sqrt{g}$. Then, the GUT symmetry is restored if
$B \agt 10^{-6}(\lambda_\Phi/\alpha)M_{\rm Pl}^2$. Using 
$B \sim (\lambda/g)^{1/2} M_{\rm Pl}^2$ we find the condition
$\alpha > \lambda_\Phi\sqrt{g}$. Combining both conditions
we see that $\eta$ particles which will satisfy $ \lambda_\Phi\sqrt{g} <
\alpha < \sqrt{g}$ (if exist) can restore the GUT symmetry.

If parametric resonance is still efficient for particles which does not satisfy 
the condition $\alpha < \sqrt{g}$, we obtain
$B \sim  m_\phi^2 M_{\rm Pl}^2/\alpha\Phi^2 \sim 10^{-7} M_{\rm Pl}^2/\alpha$. 
Symmetry restoration can be possible if $\lambda_\Phi < 10^{-1}$.

As final remarks let us discuss the applicability of our approach to the effective potential
based on Eq. (\ref{e2}). In the usual frameworks developed for equilibrium in 
Refs. \cite{1l} this would  correspond to one-loop
approximation to the effective potential. No significant further 
approximations are made despite the situation is non-equilibrium.
As compared to the equilibrium, higher loops can be important 
in the present case, however, and this issue deserves separate study.
Note also, that the effective potential is not a well defined notion in itself,
but the effective potential can be defined  for
the case of special field configurations,  and this is
sufficient for our purposes. The effective potential is equal to minus pressure
if one restricts to critical bubbles and it is equal to energy density
if one considers homogeneous field configurations only.
Equations of motion for arbitrary
field configurations can also be easily found in the present approach.
It is sufficient to use continuity of the stress-energy tensor,
$\partial_\nu T^{\mu \nu} =0$, with $T^{\mu \nu}$ being the sum of Eq. (\ref{e2})
and the stress-energy of the free field $\Phi$. With the use of the Liuville
theorem for $f(p,x)$ this continuity condition reduces to the equation
for the field $\Phi$,
\begin{equation}  
\Box \Phi + \frac{dV_0}{d\Phi} + \frac{dm_\eta^2}{d\Phi} \int \frac{d^3p}
{(2 \pi)^3} \frac{f}{2p^0} =0 \, .
\label{e10}
\end{equation}

One of the  advantages of our approach to the calculation of the effective potential 
being based on Eq. (\ref{e2}) (or Eq. (\ref{e10})) is that it allows us to find how 
the coefficient
$B$ changes when the distribution function evolves according to the kinetic equation
\cite{st95}, approaching an equilibrium.

In the present discussion we have neglected numerical factors like those which arise due 
to expansion of the Universe. Those will decrease value of $B$ somewhat. On the other
hand the usually employed description of stimulated decays based on the Mathieu equation 
takes into account only the processes of the form $n \rightarrow 2$. However, at large
phase-space density of particles when $g f(p)  > 1$ the processes $n \rightarrow m$
with  $m > 2$ start to dominate. This might reduce the evaluation of $\bar{E}$ of Ref. 
\cite{kls94} and increase the value of $B$.

We conclude that physical processes at preheating are very important, especially with 
regard to problems of symmetry restoration, and deserve detailed study.

When the first version of this work was finished I became aware of  
Ref. \cite{kls} where similar conclusions were reached. I am grateful  
to the authors of Ref. \cite{kls} for correcting some of the  
statements contained in the original version of my paper.

\acknowledgments
I thank  R. Holman, S. Khlebnikov, M. Shaposhnikov, G. Steigman and 
D. Thomas for useful discussions. 
This research was supported by the DOE grant DE-AC02-76ER01545 at Ohio.

\end{document}